\documentclass[a4paper,11pt]{article}
\pdfoutput=1 

\usepackage{jheppub} 
\usepackage{amsmath}

\usepackage{booktabs}

\usepackage{tabularx}
\usepackage{array}
\usepackage{adjustbox}

\usepackage[T1]{fontenc} 

\usepackage{mathtools}
\newcommand{\dalembertian}{\Box}

\newcommand{\bref}[1]{(\ref{#1})}
\newcommand{\dl}{\partial}
\newcommand{\gT}{\mathcal{T}} 
\newcommand{\ii}{\mathrm{i}}

\title{\boldmath Towards a Non-Perturbative Classical Double Copy}

\author[a,1]{Kymani Armstrong-Williams\note{Corresponding author.}}

\affiliation[a]{Centre for Theoretical Physics and Astronomy, School of Physical and Chemical Sciences,
Queen Mary University of London, 327 Mile End Road, London E1 4NS, UK
}

\emailAdd{k.t.k.armstrong-williams@qmul.ac.uk}

\abstract{
We construct an exact solution-generating correspondence between
restricted ansatz sectors of quartic biadjoint scalar theory, complexified
$SU(2)$ Yang--Mills theory and four-dimensional general relativity with
conformally flat metrics and traceless energy-momentum.  A factorised biadjoint field, a
Corrigan--Fairlie--'t Hooft--Wilczek gauge potential and a conformal metric are
generated by a common scalar seed.  Their equations of motion reduce to
the same equation of motion under the identification of couplings, which
retains the nonlinear dynamics on all three
legs of the correspondence.  The scalar equation captures only the trace of
the Einstein equations; deriving the trace-free part of the Einstein equation additionally yields a
source relation between Gravity and Yang-Mills. From this, we construct a non-perturbative classical double copy for the restricted ansatz sectors between quartic biadjoint scalar theory, $SU(2)$ Yang--Mills theory and gravity.
We illustrate the dictionary of mappable solutions
with rational and constant profiles representing $AdS_4$, planar $dS_4$ and
Minkowski space, and with Jacobi-elliptic profiles generating a positive-energy
recollapsing branch for negative cosmological constant and a negative-energy bouncing branch for
positive cosmological constant. Using these results, we provide a new example of the Kerr-Schild double copy.}

\begin{document} 
\maketitle
\flushbottom

\section{Introduction}
\label{sec:intro}
\begin{figure}[t]
\centering
\includegraphics[width=0.8\textwidth]{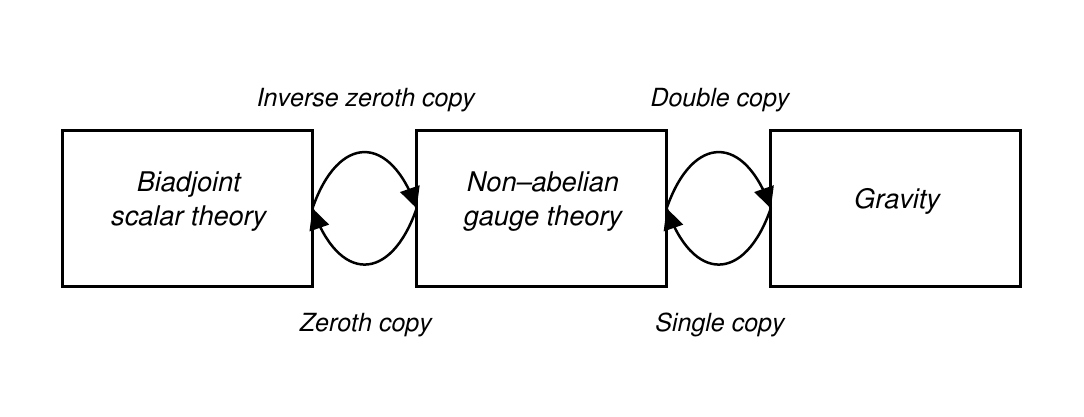}
\caption{The Classical Double Copy Framework.}
\label{fig:classicalDC}
\end{figure}

\begin{figure}[!t]
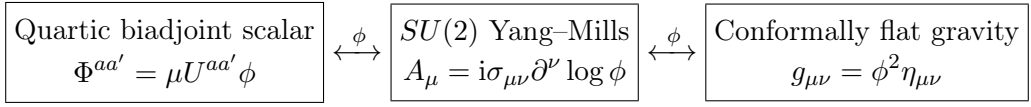

\centering
\begin{equation*}
\boxed{\begin{gathered}
\text{Quartic biadjoint scalar}\\[-1mm]
\Phi^{aa'}=\mu U^{aa'}\phi
\end{gathered}}
\xleftrightarrow{\ \phi\ }
\boxed{\begin{gathered}
SU(2)\text{ Yang--Mills}\\[-1mm]
A_\mu=\ii\sigma_{\mu\nu}\partial^\nu\log\phi
\end{gathered}}
\xleftrightarrow{\ \phi\ }
\boxed{\begin{gathered}
\text{Conformally flat gravity}\\[-1mm]
g_{\mu\nu}=\phi^2\eta_{\mu\nu}
\end{gathered}}
\end{equation*}
\caption{The common seed $\phi$ relates the three restricted ansatz sectors.} \label{fig:nonlinear-map}
\end{figure}

The double copy has become a well-known correspondence between gauge and gravity Theories both at the level of quantum scattering amplitudes and classical field theory \cite{Bern:2010ue,Bern:2008qj,Bern:2010yg}. In particular, it hints at a deep underlying connection between non-Abelian gauge theories and dravity, which to the present has not been fully realised. The double copy relationship in all its incarnations is usually made manifest through a set of well-defined replacement rules. 
\\
\\
Moreover, the double copy has also seen broad applications as a tool for calculating scattering amplitudes and observables within gravitational physics \cite{Adamo:2022dcm,Kosower_2022,Bern_2024}. Through the study of double copy, it was also realised that gauge theories can be seen to be related to an unphysical theory known as cubic biadjoint scalar theory \cite{Cachazo:2013iea,de2017extended,white2016exact,Armstrong-Williams:2026dmk,Armstrong-Williams:2025spu,Bahjat-Abbas:2018vgo,bastianelli2021worldline,de2023holonomic}; a scalar theory whose field is simultaneously valued in two Lie algebras in the adjoint representation. Cubic biadjoint scalar theory has also seen application into the study of geometrical aspects of scattering amplitudes \cite{Arkani-Hamed:2017mur,Cachazo:2013gna,Cachazo:2019ngv,Mizera:2017rqa,cachazo2020notes,frost2018biadjoint}.
Owing to its unphysical nature, where  the energy in the theory is unbounded from below, recent work has explored generalising biadjoint scalar theory via the addition of higher point interaction terms, gauging the biadjoint field or by adding mass and source terms \cite{Banerjee:2018tun,Aneesh:2019cvt,Jagadale:2020qfa,Moynihan:2021rwh,Armstrong-Williams:2025spu,Armstrong-Williams:2026dmk,Cheung:2021zvb}.\\ 
\\
The classical double copy \cite{Monteiro:2014cda,Bahjat-Abbas:2017htu,Luna:2015paa,Carrillo-Gonzalez:2017iyj,Luna:2016due,Armstrong-Williams:2022apo} describes a correspondence between classical solutions between biadjoint scalar theories, gauge theories and theories of gravity (see figure \ref{fig:classicalDC} for a schematic overview of the relationship between different theories).\footnote{A complementary nonperturbative construction exists in two-dimensional scalar theories, where the double copy can be formulated off shell at the Lagrangian level and implemented directly on classical solutions \cite{Cheung:2022mix}.}
The classical double copy was first explored in the context of Kerr-Schild solutions in General Relativity and a specific set of solutions in classical electromagnetism. This classical correspondence was then extended to Petrov type D and N gravity solutions, before extensions to $\mathcal{N}=0$ supergravity and Einstein-Maxwell gravity non-vacuum solutions \cite{Godazgar:2020zbv,White:2020sfn,Chacon:2021wbr,Luna:2018dpt,Alawadhi:2020jrv,Easson:2021asd,Luna:2022dxo,Chacon:2020fmr,Keeler:2024bdt,Keeler:2020rcv,Armstrong-Williams:2023ssz,Zhao:2024ljb,Zhao:2024wtn,Kent:2024mow,Caceres:2025eky,Kent:2025pvu,Moynihan:2025vcs,CarrilloGonzalez:2022ggn,Emond:2022uaf,Armstrong-Williams:2024bog}.\footnote{Beyond the scope of this paper is the Convolutional Double Copy, which utilises the convolution of classical fields to realise this duality \cite{Ilderton:2024oly,Anastasiou:2014qba,Anastasiou:2018rdx,Luna:2020adi,Borsten:2021zir,Godazgar:2022gfw,Liang:2023zxo}.}
\\
\\
Though very successful, the classical double copy program has been plagued by the fact that the solutions copyable under this correspondence have all been those which linearise the gravitational equations or restrict the gauge field to an effectively Abelian sector.  It therefore remains unclear how one may identify exact solutions in which nonlinear and genuinely non-Abelian field configurations can be related directly to non-linear gravitational solutions.  Conformally flat spacetimes provide a particularly sharp test: they are of Petrov type O and have vanishing Weyl tensor, so their nontrivial gravitational information lies in the Ricci tensor and matter sources rather than in the Weyl curvature used by the usual Weyl double copy.
\\
\\
In this paper, we revisit the fact that, in four dimensions, the equations of motion for SU(2) Yang-Mills theory and general relativity reduce down to the same scalar equation of motion (namely that of sourceless $\phi^{4}$ theory) under specific non-linear ans\"atze \cite{brendle2008conformal,Cadoni:2006ww,oh1979periodic,actor1979classical,oh1985nonabelian}. Previous literature on this gravitational ansatz has shown that these solutions correspond to conformally flat vacuum solutions of the Einstein equations, including the spatially flat FLRW metric. This class of solutions are all, by definition (due to the Weyl tensor vanishing) of Petrov Type O.
\\
\\
The set of SU(2) Yang-Mills solution under this ansatz, in previous literature correspond to a collection of mainly elliptic, complex valued, single parameter non-linear solutions. Even though they are complex valued, these solutions all yield real energy-momentum, much akin to light waves in classical electromagnetism. 
\\
\\
It was then shown, by direct comparison of the gravitational and Yang-Mills equation of motion \cite{sinzinkayo1984vacuum,sinzinkayo1986new,sinzinkayo1985solutions}, that the gravitational class of solutions is much larger than previously thought. In particular, it became apparent that non-vacuum solutions are allowed under this ansatz, thereby enlarging the class of gravitational
solutions to include conformally flat solutions of the
Einstein--Yang--Mills equations with traceless energy-momentum.
\\
This paper therefore re-examines these results in the context of the double copy, reinterpreting the results of refs \cite{sinzinkayo1984vacuum,sinzinkayo1986new,sinzinkayo1985solutions} as establishing a non-perturbative mapping between $SU(2)$ Yang-Mills theory and the class of conformally flat solutions of the Einstein equations with traceless energy-momentum.
More concretely, in addition to each theory reducing to an equation of motion for the same scalar field $\phi$, the energy momentum tensors for each theory are related to each other under a conformal transformation.
\\
\\
Furthermore, upon choosing the $\mathrm{SU(2)}\times\mathrm{SU(2)}$ gauge group for quartic biadjoint scalar theory, we show that, upon applying a non-linear ansatz, the equation of motion also reduces to the same scalar equation of motion as in conformally flat gravity and Yang-Mills theory.
Combining these results allows us to formulate a new non-perturbative double copy framework in which one can relate, for the first time, non-linear solutions of quartic biadjoint scalar theory, Yang-Mills theory and general relativity.
 The resulting three-way map is summarised in figure~\ref{fig:nonlinear-map}. 
 \\
 \\
The paper is organised as follows.  Sections~\ref{sec:qbast} and \ref{sec:YangMills} derive the quartic biadjoint and Yang--Mills reductions to $\phi^{4}$ theory.  Section~\ref{sec:gravity} reexamines the conformally flat gravitational sector and its extension from previous literature to spacetimes sourced by traceless energy-momentum.  Section~\ref{sec:applications} applies the correspondence to vacuum, cosmological and electrovacuum examples.  Section~\ref{sec:classical} compares the construction with the Kerr--Schild double copy, after which we conclude with a discussion of the result and its limitations.

%

\section{Quartic Biadjoint Scalar Theory}
\label{sec:qbast}

The Lagrangian density for generalised biadjoint scalar $\Phi^{a a'}$  coupled to mass $m$ and current $J^{aa'} $ with adjoint dashed and undashed indices valued in two different $SU(N)$ Lie algebras, is given by\footnote{Throughout this paper we use the signature convention $(-,+,+,+).$}

\begin{align}
\mathcal L
&= \frac{1}{2}(\partial_\mu\Phi^{aa'})(\partial^\mu\Phi^{aa'})
 - \frac{1}{2}m^2\Phi^{aa'}\Phi^{aa'}
 - \frac{y}{3}f^{abc}\tilde f^{a'b'c'}\Phi^{aa'}\Phi^{bb'}\Phi^{cc'}
\nonumber\\
&\quad
 - \frac{\lambda_4}{4}
 f^{ebc}\tilde f^{e'b'c'}
 f^{eda}\tilde f^{e'd'a'}
 \Phi^{aa'}\Phi^{bb'}\Phi^{cc'}\Phi^{dd'}
 + \Phi^{aa'}J^{aa'}, \label{eq:GBAST} 
\end{align}
\\
where $y$ and $\lambda_{4}$ are the coupling constants for the cubic and quartic interaction terms respectively, and $f^{abc}$, $\tilde{f}^{a'b'c'}$ are the SU(N) structure constants associated to each group. 
\\
In this investigation, we are only concerned with purely quartic biadjoint scalar theory, which is obtained by setting $m=y=J^{aa'}=0$. The appropriate equation of motion is then derived to be

\begin{equation}
\dalembertian\,\Phi^{aa'} \;+ \lambda_4\; \, f^{ebc}\,\tilde{f}^{e'b'c'}\, f^{eda}\,\tilde{f}^{e'd'a'}\, \Phi^{bb'}\Phi^{cc'}\Phi^{dd'} \;=\; 0 \label{eq:qbasteom}
\end{equation}
\\
The next step is then to choose an appropriate non-linear ansatz to separate the colour and spacetime information. 
We first fix the associated Lie algebra of  eq \bref{eq:qbasteom} to be that of $\mathfrak{su}(2) \times \mathfrak{su}(2)$, which induces that our structure constants become 3-D Levi Civita symbols 
\begin{equation}
f^{abc}=\varepsilon^{abc},
\qquad
\widetilde f^{a'b'c'}=\varepsilon^{a'b'c'},
\end{equation} 
\\
therefore setting eq \bref{eq:qbasteom} to be
\\
\begin{align}
\dalembertian\,\Phi^{aa'} \;+ \lambda_4\; \, \varepsilon^{ebc}\,\varepsilon^{e'b'c'}\, \varepsilon^{eda}\,\varepsilon^{e'd'a'}\, \Phi^{bb'}\Phi^{cc'}\Phi^{dd'} \;=\; 0. \label{eq:qbasteomsu2}
\end{align}
\\
We then decompose the scalar field into the product of a pure spacetime dependent field $\phi$ and a orthogonal constant $3\times3$ colour matrix $U^{aa'}$ ($U\in O(3)$)
\\
\begin{align}
\Phi^{aa'} =\mu U^{a a'} \phi , \label{eq:qbastansatz}
\end{align}
\\
where the reader will note that we have added a potentially dimensional parameter $\mu$ into the ansatz of $\Phi^{aa'}$. The inclusion of this parameter will be made apparent later, where we will wish to make use of dimensional couplings, and  in order to do so $\mu$ must take on some dimensionality, to offset that (as seen from the action) $\lambda_4$ is dimensionless.
From eq \bref{eq:qbasteomsu2}, one can then recognise that 
\\
\begin{align}
&\varepsilon^{ebc}\varepsilon^{e'b'c'}
 \varepsilon^{eda}\varepsilon^{e'd'a'}
 U^{bb'}U^{cc'}U^{dd'}=4U^{aa'},
\end{align}
\\
which when applied to the equation of motion in eq \bref{eq:qbasteom} yields 
\\
\begin{align}
\dalembertian\,\phi \;+ 4 \mu^{2}\lambda_4\; \, \phi^{3} \;=\; 0. \label{eq:qbastscalareom}
\end{align} 

\section{SU(2) Yang-Mills Theory}
\label{sec:YangMills} 

In Minkowski signature $(-,+,+,+)$, the Corrigan--Fairlie--'t Hooft--Wilczek solution \cite{oh1979periodic,actor1979classical,oh1985nonabelian,CORRIGAN197769} generates complexified solutions with real energy momentum to the SU(2) Yang-Mills equations is (given in Cartesian coordinates) by 
\begin{equation}
A_{\mu}(x) \;=\; \ii \sigma_{\mu\nu}\, \dl^{\nu} \ln(\phi), \label{eq:YMansatz}
\end{equation}
where $\sigma_{\mu\nu}$ is an antisymmetric $O(4)$ matrix, with components which are proportional to the SU(2) Pauli matrices $\sigma_{i}$, namely\footnote{Lower Case Roman indices such as $i$, $j$ and $k$ will run over value $1$ to $3$.}

\begin{equation}
\sigma_{ij} = \frac{1}{2}\epsilon_{ijk}\sigma_k,\quad
\sigma_{i0} = \frac{\ii}{2}\sigma_i,\quad
\sigma_{0i} = -\frac{\ii}{2}\sigma_i,\quad
\sigma_{\mu\mu} = 0.
\end{equation}
\\
This may be written in block form as 
\begin{equation}
\sigma_{\mu\nu} \;=\;
\begin{pmatrix}
0 & -\tfrac{\ii}{2}\sigma_1 & -\tfrac{\ii}{2}\sigma_2 & -\tfrac{\ii}{2}\sigma_3 \\
\tfrac{\ii}{2}\sigma_1 & & & \\
\tfrac{\ii}{2}\sigma_2 & \multicolumn{3}{c}{\tfrac{1}{2}\,\varepsilon^{ijk}\sigma_{k}} \\
\tfrac{\ii}{2}\sigma_3 & & &
\end{pmatrix},
\end{equation}
 \\
where the gauge field is related to the colour indexed gauge field via the following 

\begin{align}
A_{\mu} =g \frac{ \sigma^{a} }{2 \ii} A^{a}_{\mu}, \label{eq:colourgaugedef}
\end{align}
where $g$ is the Yang-Mills coupling constant.
\\
An interesting property of this class of solutions is that their corresponding vector field $A \equiv  A_{\mu}\partial^{\mu}$ generates volume-preserving diffeomorphisms

\begin{align}
\partial^{\mu} A_{\mu} = 0, \label{eq:areadiffcon}
\end{align}
\\
which follows from the antisymmetry of $\sigma_{\mu \nu}$. 
\\
\\
The vector field generated by $A_{\mu}$ is then given by 
\\
\begin{align}
A_{\mu}\dl^{\mu}
&=\frac{1}{2}
\Bigl[
-\sigma_i\,(\partial_i\ln(\phi))\partial_0
+
\sigma_i\,(\partial_0\ln(\phi))\partial_i
+
\ii\epsilon_{ijk}\sigma_k\,
\partial_j\ln\phi\,\partial_i
\Bigr]. \label{eq:ymvectorfield}
\end{align}
\\
Defining the two vectors $\alpha_i$ and $\beta_i$ by 
\\
\begin{align}
\alpha_{i} \;\equiv\; \dl_{i}, \qquad \beta_{i} \;\equiv\; \dl_{i}\ln(\phi),
\end{align}
\\
we can rewrite eq \bref{eq:ymvectorfield} in the form
\\
\begin{align}
A_{\mu}\dl^{\mu}
&= \tfrac{1}{2}\Bigl[\,-\sigma_{i}\,\beta^{i}\,\dl_{0} +\dl_{t}\ln(\phi)\,\sigma_{i}\,\alpha^{i} + \ii \varepsilon^{ijk}\sigma_{k}\,\beta^{j}\,\alpha^{i}\Bigr]. \label{eq:ymvectorfieldrelabelled}
\end{align}
\\
Equation \bref{eq:ymvectorfieldrelabelled} is then interpreted as a vector field which generates diffeomorphisms, inducing non-trivial mixing between colour and spacetime. Structures such as those seen in eq \eqref{eq:ymvectorfieldrelabelled} may provide pathways to explore novel geometrical interpretations (including relationships to so-called kinematic algebras \cite{Monteiro:2011pc,Monteiro:2013rya,Reiterer:2019dys,Borsten:2022vtg,Borsten:2021hua,Bonezzi:2023pox,Bonezzi:2026qxf,Armstrong-Williams:2024icu}), and thus merit further study. 
\\
Upon substitution eq \bref{eq:YMansatz} into the SU(2) Yang -Mills equations, the resulting equation of motion is
\\
\begin{align}
\Box \phi + \lambda \phi^{3} = 0, \label{eq:YMscalarequation}
\end{align}
\\
where $\lambda$ is some arbitrary coefficient of integration with mass dimension 2. Non-linear solutions are produced upon the choice 
\\
\begin{align}
\lambda \neq 0,
\end{align}
\\
whereas the choice $\lambda = 0$ reduces eq \bref{eq:YMscalarequation} to 
\\
\begin{align}
\Box \phi = 0. \label{eq:YMSD}
\end{align}
\\
This corresponds to self-dual variants of eq. \bref{eq:YMansatz} that
satisfy the $SU(2)$ Yang--Mills equations.
\\
Crucially, we can directly identity the form of eq \eqref{eq:YMscalarequation} with the equation of motion for Quartic Biadjoint scalar theory in eq \bref{eq:qbastscalareom}, up to a replacement of coupling 

\begin{align}
4\mu^{2} \lambda_{4} \rightarrow \lambda. \label{eq:couplingqbastym}
\end{align}
\\
Equation \bref{eq:couplingqbastym} then allows us to make a direct non-perturbative map between sub-sectors of quartic biadjoint scalar theory and Yang-Mills theory, and determines that $\mu$ has mass dimension $1$.\\
Interestingly, solutions to eq \bref{eq:YMscalarequation} have a real
energy-momentum tensor
\begin{align}
T_{\mu\nu}(x) =& \frac{\lambda}{g^{2}}\left[4\,\partial_{\mu}\phi\,\partial_{\nu}\phi - 2\phi\,\partial_{\mu}\partial_{\nu}\phi - \eta_{\mu\nu}\left (\frac{\lambda\phi^{4}}{2} + \partial_{\alpha}\phi\,\partial^{\alpha}\phi \right)\right] \notag \\ 
=& \frac{\lambda}{g^{2}}\left[4\,\partial_{\mu}\phi\,\partial_{\nu}\phi - 2\phi\,\partial_{\mu}\partial_{\nu}\phi - \eta_{\mu\nu}\left (\frac{\lambda\phi^{4}}{2} + (\partial \phi)^{2} \right)\right],
\notag \\ 
\equiv& \frac{\lambda}{g^{2}} \left [4\,\partial_{\mu}\phi\,\partial_{\nu}\phi
-2\phi\,\partial_{\mu}\partial_{\nu}\phi
-\eta_{\mu\nu}
\left(
-\frac12\phi\Box\phi 
+(\partial\phi)^{2}
\right)\right ],
\label{eq:YMansatzEnergyMomemtum}
\end{align}
where we have defined
\begin{align}
(\partial \phi)^{2}\equiv \partial_{\alpha}\phi\,\partial^{\alpha}\phi.
\end{align}
\\
The class of solutions generated by eq \bref{eq:YMansatz} (as described in refs \cite{oh1985nonabelian,oh1979periodic}) is best understood by first making a change of variables to a new variable $u$ in eq \bref{eq:YMscalarequation} (see refs \cite{Frasca:2009bc,Armstrong-Williams:2025spu} for similar ideas)
\begin{align}
u = p_{\mu}x^{\mu} + q, \label{eq:changeofv}
\end{align}
\\
where $p_{\mu}$ is a generic constant 4-vector and $q$ is a constant. Equation \bref{eq:YMscalarequation} then becomes 
\\
\begin{align}
p^{2} \frac{d^{2} \phi (u)}{d u^{2}} + \lambda \phi^{3}(u) = 0. \label{eq:YMScalarUnull}
\end{align} 
\\
If $ p^{2} \neq 0$, we may multiple both sides of eq \bref{eq:YMScalarUnull} by $p^{-2}$
\\
\begin{align}
\frac{d^{2} \phi (u)}{d u^{2}} + \frac{\lambda}{p^{2}} \phi^{3}(u) = 0. \label{eq:YMScalarU}
\end{align} 
\\
Applying the single-parameter restriction $\phi\equiv\phi(u)$ to the
initial gauge-field ansatz in eq \bref{eq:YMansatz} yields

\begin{align} 
A_{\mu}(x) \;=\; \ii \sigma_{\mu\nu}\, p^{\nu} \frac{\phi'(u)}{\phi(u)} \label{eq:singlevarYMansatz}
\end{align} 
\\
where $\phi'(u) \equiv \frac{\partial \phi (u)}{\partial u}$.
The solutions to eq \bref{eq:singlevarYMansatz} are outlined in table \ref{tab:ym-solutions}. Crucially, the form that eqs \eqref{eq:YMScalarUnull} and \eqref{eq:YMScalarU} take can be translated directly into quartic biadjoint scalar theory via the nonlinear
relationship defined in eq. \bref{eq:couplingqbastym}. Consequently, every solution in table \ref{tab:ym-solutions} has a direct counterpart in Quartic Biadjoint scalar theory. 
\\
\\
In the next section, we consider a class of gravitational
solutions---conformally flat spacetimes with traceless
energy-momentum---which we relate directly, through a nonlinear
correspondence, to the solutions in table \ref{tab:ym-solutions} and hence
to quartic biadjoint scalar theory.

\begin{table}[htbp]
\centering
\renewcommand{\arraystretch}{1.35}
\setlength{\tabcolsep}{5pt}
\begin{adjustbox}{max width=\textwidth}
\begin{tabular}{@{}cccc@{}}
\toprule
\(\phi\)
& Condition on \(p_\mu\)
& Real zeros/poles
& Energy--momentum \\
\midrule

\(\displaystyle \phi_1(x)=\frac{1}{p\cdot x+q}\)
&
\(\displaystyle 2p^2=-\lambda\)
&
\(\displaystyle -\)
&
\(\displaystyle T_{\mu\nu}=0\)
\\

\midrule

\(\displaystyle \phi_2(x)=\operatorname{cn}(u,k)\)
&
\(\displaystyle p^2=\lambda\)
&
Zeroes at \(\displaystyle u=(2N+1)K,\; N\in\mathbb{Z};\; \text{period }4K\text{ in }u\)
&
\(\displaystyle T_{\mu\nu}
=\frac{\lambda}{2g^2}
\left(4p_\mu p_\nu-\eta_{\mu\nu}p^2\right)\)
\\

\midrule

\(\displaystyle \phi_3(x)=\operatorname{nc}(u,k)\)
&
\(\displaystyle p^2=-\lambda\)
&
Poles at \(\displaystyle u=(2N+1)K,\; N\in\mathbb{Z};\; \text{period }4K\text{ in }u\)
&
\(\displaystyle T_{\mu\nu}
=-\frac{\lambda}{2g^2}
\left(4p_\mu p_\nu-\eta_{\mu\nu}p^2\right)\)
\\

\midrule

\(\displaystyle \phi_4(x)=\operatorname{sd}(u,k)\)
&
\(\displaystyle p^2=2\lambda\)
&
Zeroes at \(\displaystyle u=2NK,\; N\in\mathbb{Z};\; \text{period }4K\text{ in }u\)
&
\(\displaystyle T_{\mu\nu}
=\frac{\lambda}{g^2}
\left(4p_\mu p_\nu-\eta_{\mu\nu}p^2\right)\)
\\

\midrule

\(\displaystyle \phi_5(x)=\operatorname{ds}(u,k)\)
&
\(\displaystyle p^2=-\frac{\lambda}{2}\)
&
Poles at \(\displaystyle u=2NK,\; N\in\mathbb{Z};\; \text{period }4K\text{ in }u\)
&
\(\displaystyle T_{\mu\nu}
=-\frac{\lambda}{4g^2}
\left(4p_\mu p_\nu-\eta_{\mu\nu}p^2\right)\)
\\

\bottomrule
\end{tabular}
\end{adjustbox}
\caption{Plane-wave-type solutions of the reduced Yang--Mills equation. Here \(u=p\cdot x+q\); \(p^2=p_\mu p^\mu\); \(k^2=1/2\); \(K\simeq 1.854\) is the complete elliptic integral of the first kind, and \(N\in\mathbb{Z}\).}
\label{tab:ym-solutions}
\end{table}

\section{General Relativity and Conformally Flat Solutions}
\label{sec:gravity}
Recall the Einstein equations with non-vanishing cosmological constant $\Lambda$ (with Newton's constant $G_{N}$) are given by
\\
\begin{align}
R_{\mu \nu} - \frac{1}{2} R g_{\mu \nu} + \Lambda g_{\mu \nu}  = 8 \pi G_{N} \gT_{\mu \nu}. \label{equ:efe}
\end{align} 
\\
Making the ansatz that our matter is conformally invariant ($\gT^{\mu}_{\mu} =0$), then taking the trace of eq \bref{equ:efe} --- contracting it with
$g^{\mu\nu}$---yields \cite{forger2004currents}
\\
\begin{align}
R = 4 \Lambda. \label{equ:efetr}
\end{align}
\\\
The class of conformally flat solutions to the vacuum Einstein equations can be represented by a metric of the form
\\
\begin{align}
g_{\mu \nu} = \phi^{2} \eta_{\mu \nu}, \label{eq:conformalmetric}
\end{align}
\\
when applied to eq \bref{equ:efetr} for any choice of $\phi$ yields, 
\\
\begin{align}
\frac{-6 \Box \phi}{\phi^{3}}= 4 \Lambda, \label{equ:efetrconformal}
\end{align} 
which when rearranged yields
\\
\begin{align}
\Box \phi +\frac{2}{3} \Lambda \phi^{3} = 0.  \label{eq:grscalarequ}
\end{align} 
\\
Notice that eq \bref{eq:grscalarequ} has the same form as the equations
for quartic biadjoint scalar theory and Yang--Mills theory in eqs
\bref{eq:qbastscalareom} and \bref{eq:YMscalarequation}, respectively,
under the following replacements of the coupling multiplying the
interaction term (noting also that the cosmological constant has mass
dimension $2$)

\begin{align}
4 \mu^{2} \lambda_{4} \rightarrow \lambda \rightarrow \frac{2}{3}\Lambda. \label{eq:coupgrymb}
\end{align}
Equation \bref{eq:coupgrymb} demonstrates that there exists a non-perturbative relationship between solutions of quartic biadjoint scalar theory, Yang Mills theory and gravity.In particular, it allows us to identify gravitational counterparts of the Yang--Mills solutions in table \ref{tab:ym-solutions}\\
\\
The Ricci tensor and Ricci scalar for eq \bref{eq:conformalmetric} are given by 

%

\begin{align}
&R_{\mu\nu}
= -\frac{2}{\phi}\,\partial_\mu \partial_\nu \phi
+ \frac{4}{\phi^2}\,\partial_\mu \phi\,\partial_\nu \phi
- \eta_{\mu\nu}
\left[
\frac{\Box \phi}{\phi}
+ \frac{(\partial \phi)^2}{\phi^2}
\right],  \label{eq:conformalmetricricci}\\ 
&R = \frac{-6 \Box \phi}{\phi^{3}}. \label{eq:conformalmetricscalar}
\end{align}
\\
Comparing eq \bref{eq:grscalarequ} with \bref{eq:YMScalarU}, we see that eq \bref{eq:grscalarequ} reduces  to an equation in the single
variable $u$
\\
\begin{align}
\frac{d^{2}\phi(u)}{du^{2}} +\frac{2}{3} \frac{\Lambda}{p^{2}} \phi(u)^{3} = 0, \label{eq:grscalarequU}
\end{align} 
from which we can take a first integral, obtaining the constant of motion $\mathfrak{C}$
\begin{align}
p^{2} \left(\frac{d\phi(u)}{du}\right)^{2}+ \frac{\Lambda}{3} \phi^{4} = \mathfrak{C}. \label{eq:grscalarfirstintU}
\end{align}
\\
Seemingly, equation \bref{eq:grscalarequ} is not enough to completely fix the dynamics of this ansatz. Namely, one also needs to consider the trace free part of the Einstein equations \cite{carroll2019spacetime}. Recalling that any symmetric rank 2 tensor $X_{\mu \nu}$ can be decomposed into its trace free parts through

\begin{align}
X^{\mathrm{TF}}_{\mu\nu}=X_{\mu\nu}-\frac{1}{4} g_{\mu\nu} X^{\rho}{}_{\rho}, \label{eq:xtracefree}
\end{align}
\\
we can apply eq \bref{eq:xtracefree} to both sides of eq \bref{equ:efe}, which yields
\\
\begin{align}
R_{\mu\nu} - \frac{1}{4} R g_{\mu\nu} = 8 \pi G_{N} \gT_{\mu \nu}. \label{eq:rtracefree}
\end{align}
\\
Upon applying eqs. \eqref{eq:conformalmetricricci} and
\eqref{eq:conformalmetricscalar}, the left-hand side of eq.
\bref{eq:rtracefree} reduces to

\begin{align}
R_{\mu\nu} - \frac{1}{4} R g_{\mu\nu}
=
-\frac{2}{\phi}\,\partial_\mu \partial_\nu \phi
+\frac{4}{\phi^2}\,\partial_\mu \phi\,\partial_\nu \phi
+\eta_{\mu\nu}
\left[
\frac{\Box \phi}{2\phi}
-\frac{(\partial \phi)^2}{\phi^2}
\right], \label{eq:tracelessgrLHS}
\end{align} 
\\
application of eq \bref{eq:grscalarequ} then yields
\\
\begin{align}
R_{\mu\nu} - \frac{1}{4} R g_{\mu\nu}
=
-\frac{2}{\phi}\,\partial_\mu \partial_\nu \phi
+\frac{4}{\phi^2}\,\partial_\mu \phi\,\partial_\nu \phi
+\eta_{\mu\nu}
\left[
-\frac{1}{3}\Lambda  \phi^{2} 
-\frac{(\partial \phi)^2}{\phi^2}
\right], \label{eq:tracelessgrLHS2}
\end{align}
\\
meaning eq \bref{eq:rtracefree} in its entirety becomes the definition for the traceless energy-momentum  $\gT_{\mu \nu}$ 

\begin{align}
\gT_{\mu \nu}
= \frac{1}{8 \pi G_{N}}
\left( -\frac{2}{\phi}\,\partial_\mu \partial_\nu \phi
+\frac{4}{\phi^2}\,\partial_\mu \phi\,\partial_\nu \phi
+\eta_{\mu\nu}
\left[
-\frac{1}{3}\Lambda  \phi^{2} 
-\frac{(\partial \phi)^2}{\phi^2}
\right]\right), \label{eq:emgrtraceless}
\end{align}
and for $\phi \equiv \phi(u)$ becomes (under the application of eqs \bref{eq:grscalarequU} and \bref{eq:grscalarfirstintU} to eq \bref{eq:emgrtraceless})
\begin{align}
\mathcal T_{\mu\nu}
=
\frac{\mathfrak C}{8\pi G_N\,\phi^2}
\left(
\frac{4p_\mu p_\nu}{p^2}-\eta_{\mu\nu}
\right). \label{eq:emgrtracelessU}
\end{align}
\\
Comparing the energy-momentum tensor for the class of $SU(2)$ Yang--Mills
solutions in eq \bref{eq:YMansatzEnergyMomemtum} with that derived for
the conformally flat solutions in eq.\bref{eq:emgrtraceless}, we see that
the energy-momentum tensors in the two theories are related by the
rescaling
\\
\begin{align}
\gT_{\mu \nu} \rightarrow \frac{g^{2}}{8 \pi G_{N} \lambda} \phi^{-2} T_{\mu \nu}. \label{eq:ymgrem0}
\end{align}
\\
Recall that, in Yang-Mills theory, the generic energy momentum tensor $T^{YM}_{\mu \nu}$ is defined as (with the $\operatorname{Tr}$ operator acting on the colour indices rather than the spacetime ones)
\begin{align}
T^{YM}_{\mu\nu}
=
-\frac{2}{g^{2}}
\operatorname{Tr}\left(
F_{\mu\rho}F_{\nu}{}^{\rho}
-\frac14 \eta_{\mu\nu}F_{\rho\sigma}F^{\rho\sigma}
\right). \label{eq:OGYMEM}
\end{align} 
\\
Taking into consideration that the specific energy-momentum tensor for the
class of $SU(2)$ Yang--Mills solutions in eq.
\eqref{eq:YMansatzEnergyMomemtum} must equal the expression in eq.
\eqref{eq:OGYMEM}, we can rewrite eq. \bref{eq:ymgrem0} as

\begin{align}
\gT_{\mu \nu} \rightarrow \frac{-2}{8 \pi G_{N} \lambda} \phi^{-2} \operatorname{Tr}\left(
F_{\mu\rho}F_{\nu}{}^{\rho}
-\frac14 \eta_{\mu\nu}F_{\rho\sigma}F^{\rho\sigma}
\right) . \label{eq:ymgrem}
\end{align}
\\
Thus, the gravitational energy-momentum tensor for the class of
conformally flat metrics with conformally invariant matter in eq.
\bref{eq:emgrtraceless} can be interpreted as a conformal rescaling by
$\phi^{-2}$ of the energy-momentum tensor for the class of Yang--Mills
solutions in eq. \bref{eq:YMansatzEnergyMomemtum}.
Therefore, substituting eq \eqref{eq:ymgrem} into the Einstein equations in eq \bref{equ:efe} yields the relationship

\begin{align}
R_{\mu \nu} - \frac{1}{2} R g_{\mu \nu} + \Lambda g_{\mu \nu}  = \frac{-2}{\phi^{2}\lambda} \operatorname{Tr}\left(
F_{\mu\rho}F_{\nu}{}^{\rho}
-\frac14 \eta_{\mu\nu}F_{\rho\sigma}F^{\rho\sigma}
\right). \label{equ:efe2}
\end{align} 
\\
We can think of eq \eqref{equ:efe2} as itself as a more canonical representation of the classical double copy. In order to make this manifest, it is useful to encapsulate the LHS of eq \eqref{equ:efe2} as some tensor $K_{\mu \nu}$

\begin{align}
K_{\mu \nu} = R_{\mu \nu} - \frac{1}{2} R g_{\mu \nu} + \Lambda g_{\mu \nu},  \label{eq:kdef}
\end{align} 
\\
which ultimately reduces the form of eq \eqref{equ:efe2}, yielding

\begin{align}
K_{\mu \nu} = 
-\frac{2}{\phi^{2} \lambda} \operatorname{Tr}\left(
F_{\mu\rho}F_{\nu}{}^{\rho}
-\frac14 \eta_{\mu\nu}F_{\rho\sigma}F^{\rho\sigma}
\right). \label{eq:NLDC0}
\end{align} 
\\
Importantly, although a factor of $\lambda^{-1}$ is present in eq.
\bref{eq:NLDC0}, the energy-momentum tensor evaluated for the specific
$SU(2)$ Yang--Mills ansatz also yields an overall factor of $\lambda$,
which cancels the $\lambda^{-1}$ term and eliminates any explicit
dependence on $\lambda$. To make eq. \bref{eq:NLDC0} more concrete, define
$Q_{\mu\nu}$ as the rescaled energy-momentum tensor
\begin{align}
Q_{\mu \nu} &= \frac{-2}{\lambda} \operatorname{Tr}\left(
F_{\mu\rho}F_{\nu}{}^{\rho}
-\frac14 \eta_{\mu\nu}F_{\rho\sigma}F^{\rho\sigma}
\right) \notag \\
&\equiv -2 \operatorname{Tr}\left(
\tilde{F}_{\mu\rho}\tilde{F}_{\nu}{}^{\rho}
-\frac14 \eta_{\mu\nu}\tilde{F}_{\rho\sigma}\tilde{F}^{\rho\sigma}
\right) \label{eq:Qdef}
\end{align}  
\\
where $\tilde{F}_{\mu\nu}$ is the rescaled field-strength tensor obtained
by absorbing the factor of $\lambda^{-1}$ into the energy-momentum tensor
in eq. \bref{eq:Qdef}. This removes all explicit dependence on $\lambda$
when $\tilde{F}_{\mu\nu}$ is evaluated for the specific $SU(2)$
Yang--Mills ansatz.
Therefore, when eq \bref{eq:Qdef} is applied to eq \bref{eq:NLDC0} it yields

\begin{align}
K_{\mu \nu} = 
\frac{1}{\phi^{2} } Q_{\mu \nu}. \label{eq:NLDC}
\end{align} 
\\
Structurally, eq. \eqref{eq:NLDC} has the form expected in the classical
double copy (as also seen in refs.
\cite{Monteiro:2021ztt,Armstrong-Williams:2024bog}). The left-hand side
contains a gravitational tensor describing the curvature of spacetime. The
numerator on the right-hand side contains the square of a field-strength
tensor, divided by $\phi^2$ in the denominator.
\\
However, eq \eqref{eq:NLDC} differs significantly from the previous approach taken in refs \cite{Monteiro:2021ztt,Armstrong-Williams:2024bog}. Firstly, the double copy relationship here is non-perturbative on both sides of the equation, relating a non-linear gravity solution with a non-linear SU(2) Yang-Mills solution. Secondly, the scalar field in the denominator of eq \eqref{eq:NLDC} is a solution to Quartic biadjoint scalar theory instead of the standard cubic biadjoint scalar theory. Another contrast is that the biadjoint scalar is squared in this instance as opposed to the usual case where the scalar field is left unsquared in the denominator.
\\
\\
To reiterate, we have demonstrated that the four dimensional Einstein equations for conformally flat metrics with traceless energy momentum tensor reduces down to two equations: a trace equation \bref{eq:grscalarequ} and a trace-free equation \bref{eq:emgrtraceless}. 
We identity the trace equation, up to a change of coupling constant, with the same reduced equation of motion obtained for Yang-Mills theory in eq \bref{eq:YMscalarequation} and Quartic Biadjoint scalar theory in eq \bref{eq:qbastscalareom}. This allows us to map the one-parameter
Yang--Mills solutions in table \ref{tab:ym-solutions} directly to gravity.
The trace-free equation then provides a direct mapping between the energy momentum tensor of the  
gravitational and Yang-Mills solutions. Specifically, for the class of solutions considered here, the tensors are
related by the conformal transformation in eq \bref{eq:ymgrem}.
Finally, this observation yields a non-perturbative classical double copy relationship between conformally flat solutions of the Einstein equations with traceless energy momentum and the class of SU(2) Yang-Mills equations defined in eq \bref{eq:YMscalarequation}.
\\
\\
In section \ref{sec:applications}, we explore examples of the types of solutions which are related through this correspondence.

\section{Applications} 

\label{sec:applications}

In this section, we explore matching up SU(2) Yang-Mills solutions in Table \ref{tab:ym-solutions} to canonical counterparts in general relativity. We begin by exploring the vacuum and non-vacuum one parameter solutions discussed in Table \ref{tab:ym-solutions} before extending our discussion to solutions outside this original sector of solutions, namely those found in references \cite{sinzinkayo1984vacuum,sinzinkayo1986new,sinzinkayo1985solutions}.

\subsection{Vacuum Non-linear Solutions} 

From $\phi_{1}$ given in Table \ref{tab:ym-solutions}, namely 

\begin{align}
\phi_{1} = \frac{1}{p \cdot x +q}, 
\end{align} 
\\
The profile $\phi_1$ in table \ref{tab:ym-solutions} gives a class of
solutions corresponding to spacetimes with $\gT_{\mu\nu}=0$.
Three examples which belong to this class, depending on the sign of $\Lambda$ are

\begin{enumerate}
\item $\Lambda < 0$ : Pure $AdS_4$ in the Poincar\'e patch  
\item $\Lambda > 0$: de Sitter space ($dS_{4}$) in Planar Coordinates 
\cite{Spradlin:2001pw,Galante:2023uyf}
\item Minkowski Space for $\Lambda =0$
\end{enumerate} 
Starting with pure $AdS_{4}$, the line element for the Poincar\'e patch is given by 
\begin{align}
ds^{2} = \frac{L^{2}}{z^{2}} \left(-dt^{2} +dx^{2}+dy^{2}+dz^{2}\right), \label{eq:Ads4metric}
\end{align}
\\
where $L$ is the constant $AdS_{4}$ radius and $z >0$.
In order for eq \bref{eq:Ads4metric} to satisfy eq \bref{eq:grscalarequ}, $\Lambda$ is fixed to take the value 
\begin{align}
\Lambda = \frac{-3}{L^{2}}.
\end{align}
We can identify the scalar field $\phi$ from eq \bref{eq:Ads4metric} to be  
\\
\begin{align}
\phi = \frac{L}{z}. \label{eq:ads4phi}
\end{align}
\\
Substituting eq \bref{eq:ads4phi} into the $SU(2)$ Yang--Mills ansatz
yields
\\
\begin{align}
A_{\mu} = - \ii \sigma_{\mu 3} \frac{1}{z}, \label{eq:adsgauge}
\end{align}
\\
and, similarly, the Quartic Biadjoint scalar ansatz 
\\
\begin{align}
\Phi^{a a'} = \mu U^{a a'} \frac{L}{z}, \label{eq:adsbast}
\end{align}
\\
thereby fixing the coupling constant $\lambda_{4}$ to be 
\\
\begin{align}
\lambda_{4} = \frac{-1}{2 L^{2}\mu^{2}}.
\end{align}
\\
Moving to the de Sitter case, its line element is given in the form (for $-\infty < \eta < 0$)
\\
\begin{align}
ds^{2} = \frac{1}{H^{2} \eta^{2}} \left[ -d\eta^{2} + dx^{2} +dy^{2}+dz^{2} \right ], \label{eq:flatdsmetric}
\end{align}
where the value of $H$ is fixed by the constraint (in order to satisfy eq \bref{eq:grscalarequ})
\begin{align}
H^{2} = \frac{\Lambda}{3}.
\end{align}
The scalar field $\phi$ can then be identified as 
\begin{align}
\phi = -\frac{1}{H \eta}, \label{eq:dsphi}
\end{align}
\\
where upon substitution into \bref{eq:YMansatz} gives 
\\
\begin{align}
A_{\mu} =  \ii \sigma_{\mu 0} \frac{1}{\eta}
\end{align}
\\
and further insertion into \bref{eq:qbastansatz} yields 
\\
\begin{align}
\Phi^{a a'} = \frac{-\mu}{H\eta} U^{a a'},
\end{align}
\\
fixing the value of $\lambda_{4}$ to be 
\\
\begin{align}
\lambda_4 = \frac{\Lambda}{6\mu^{2}}.
\end{align}
\\
Finally, Minkowski space is also included in this construction, by choosing that first $\Lambda = 0$ (and by consequence $\lambda = \lambda_{4} = 0$), 
eq \bref{eq:grscalarequ} becomes 

\begin{align}
\Box \phi = 0\label{eq:grscalarequ0}
\end{align} 
\\
then choosing that our scalar function $\phi$ is a constant value $\mathfrak{s}$, eq \bref{eq:conformalmetric} becomes 

\begin{align}
g_{\mu \nu} = \mathfrak{s}^{2} \eta_{\mu\nu}.
\end{align}
\\
Consequently, this ensures that our ans\"atze for SU(2) Yang-Mills and biadjoint scalar theory become 

\begin{align}
A_{\mu} &= \ii\sigma_{\mu\nu}\partial^{\nu} \mathrm{ln}(\mathfrak{s}) \notag \\
&= 0,\label{eq:mingaugecounter}
\end{align}
\begin{align}
\Phi^{aa'} = \mu U^{aa'} \mathfrak{s}.
\end{align} 
Therefore, showing the counterpart of Minkowski metric in gauge theory is a trivial gauge field; whereas the corresponding configuration in quartic biadjoint scalar theory is a constant scalar field profile filling all space. 
Equation \bref{eq:mingaugecounter} therefore re-derives the previously known result in the literature from a new non-linear perspective, that the corresponding structure for Minkowski space in gauge theory is indeed the trivial gauge field \cite{Bahjat-Abbas:2017htu}.

\subsection{Elliptic Scalar Functions}

Table \ref{tab:ym-solutions} (given by $\phi_2$ to $\phi_5$) also discusses elliptic functions as seeds for non-linear solutions in SU(2) Yang-Mills. These elliptic solutions when translated into gravitational solutions fall into two separate classes related to the spatially flat FLRW metrics in gravity on the provision that $u$ is solely a function of time $\eta$ \cite{Galtsov:2010yqh,galt1991yang,cervero1978classical}
\\
\begin{enumerate}
\item $\phi_{2}$ and $\phi_{4}$: $\{\operatorname{cn},\operatorname{sd}\}
\quad\longleftrightarrow\quad
\text{positive-radiation FLRW with }\Lambda<0, $
\item $\phi_{3}$ and $\phi_{5}$: $\{\operatorname{nc},\operatorname{ds}\}
\quad\longleftrightarrow\quad
\text{negative-radiation bouncing FLRW with }\Lambda>0.$
\end{enumerate}
To illustrate this point succinctly, we chose the simplest parametrisation for $u$ in coordinates $\left(\eta,x,y,z\right)$, for $\omega$ as some constant parameter
\begin{align}
u = \omega \eta, \label{eq:simpparm}
\end{align}
\\
For the first case, e.g., $\phi\equiv\operatorname{cn}$, table
\ref{tab:ym-solutions} fixes the value of $\omega$ as

\begin{align}
 \omega = \sqrt{-\frac{2 \Lambda}{3}}.
\end{align}
From eq \bref{eq:conformalmetric} we can construct the metric 
\begin{align}
ds^{2} = \mathrm{cn^{2}}\left(\omega \eta,\frac{1}{\sqrt{2}}\right )\left[-d\eta^{2}+dx^{2}+dy^{2}+dz^{2}\right],\label{eq:cnmetric}
\end{align}
with energy momentum tensor (via the introduction of two parameters $a$ and $b$ which are yet to be defined)
\begin{align}
\mathcal{T}_{\mu \nu} = 
\begin{pmatrix}
a&0&0&0 \\ 
0&b&0&0 \\ 
0&0&b&0 \\ 
0&0&0&b\\ 
\end{pmatrix}, \label{eq:cntm}
\end{align}
\\
The trace of eq \bref{eq:cntm} vanishes provided that 
\\
\begin{align}
b= \frac{a}{3}, \label{eq:abrad}
\end{align}
\\
Upon closer inspection, the form of eq \bref{eq:cntm} is similar to the expected expression for a perfect fluid observed in its rest frame
\\
\begin{align}
T_{\mu\nu}
=(\rho_{0}+\mathcal{P}_{0})u_\mu u_\nu+\mathcal{P}_{0}\,g_{\mu\nu}, \label{eq:energymomfluid}
\end{align}
\\
here $\rho_0$ is the energy density, $\mathcal{P}_0$ is pressure and $u_{\mu}$ is 4-velocity, all measured in the rest frame. 
Rescaling $u^{\mu}$ via

\begin{align}
\tilde{u}^{\mu} = \phi^{-1} u^{\mu}, \label{eq:unorm}
\end{align}
\\
such that it obeys the condition 

\begin{align}
g_{\mu\nu}\tilde{u}^{\mu} \tilde{u}^{\nu} =-1,
\end{align}
\\
we can therefore derive the proper expression for $\rho_{0}$ by contracting eq \bref{eq:cntm} with $\tilde{u}^{\mu}$
\begin{align}
\rho_{0} &= \mathcal{T}_{\mu \nu} \tilde{u}^{\mu} \tilde{u}^{\nu} \notag \\
&= a \phi^{-2}, \label{eq:rhorescale}
\end{align}
which allows us to amend eq \bref{eq:abrad} to 
\begin{align}
\mathcal{P}_{0} = \frac{\rho_{0}}{3}. \label{eq:abrad2}
\end{align}
\\
In the context of the FLRW metrics, eq \bref{eq:abrad2} corresponds to a radiation-dominated universe.
\\
Comparing eq \bref{eq:cntm} with eq \bref{eq:emgrtracelessU} allows us to derive 

\begin{align}
\rho_{0}=
-\frac{3\mathfrak C}{8\pi G_N \phi^4}.  \label{eq:cndefrho}
\end{align}
\\
For the specific spacetime in eq. \bref{eq:cnmetric}, the constant of
motion in eq. \bref{eq:grscalarfirstintU} is
\begin{align}
\mathfrak{C} = \frac{\Lambda}{3},
\end{align}
\\
which sets $\rho_0$ to 
\\
\begin{align}
\rho_{0} = -\frac{\Lambda}{8\pi G_N \left(\mathrm{cn}\left(\omega \eta,\frac{1}{\sqrt{2}}\right )\right)^4},  \label{eq:cndefrhoval}
\end{align} 
\\
which for $\Lambda < 0$ is always positive. 
The square of the scale factor $\phi^{2}(t)$ for eq \bref{eq:cnmetric} in cosmic time (as a coordinate transformation away from conformal time $dt= \phi d\eta$) is given by 
\begin{align}
\phi^2(t)
=\sin\left[
2\sqrt{\frac{|\Lambda|}{3}}
(t-t_{\rm b})
\right], \label{eq:cncosmic}
\end{align}
where $t_{b}$ is the cosmic time at the point of the big bang. Recalling that the scale factor measures the size of the universe since the big bang (e.g $\phi(t) = 0$ would correspond to the universe at the point of the big bang), then the implication of eq \bref{eq:cncosmic} is that this spacetime describes a universe which expands up to maximum expansion in some time, before contracting to again back to the initial size at the point of the big bang.
The time $t_{max}$ taken for the universe to reach maximum expansion is 
\begin{align}
t_{max} =\frac{1}{4} \sqrt{\frac{3}{\lvert \Lambda \rvert} }\pi + t_b.
\end{align}
The elapsed time from the big bang until complete recollapse,
$t_{\mathrm{contract}}$, is
\begin{align}
t_{contract} =\frac{1}{2} \sqrt{\frac{3}{\lvert \Lambda \rvert} }\pi + t_b.
\end{align}
\\
The results above for the choice $\phi_{2}$ are related to the results that would follow from the choice of $\phi_{4} = \mathrm{sd}$, once one recalls the identity
\\
\begin{align}
\mathrm{sd}(u,k)
= -\sqrt2\,\mathrm{cn}(u+K,k),
\end{align}
\\
and that $\omega$ is transformed via (as dictated by Table \ref{tab:ym-solutions})

\begin{align}
\omega \rightarrow \sqrt{2} \omega.
\end{align}
\\
This demonstrates, at least in spirit, that one can obtain (up to a common phase and scaling) the result for $\phi_{4}$ by numerically adjusting $\phi_2$.
\\
\\
Moving on to the negative radiation case with $\Lambda > 0$,  we choose $\phi = \mathrm{nc}$, giving the line element 
\\
\begin{align}
ds^2=
\operatorname{nc}^2\!\left(
\sqrt{\frac{2\Lambda}{3}}\,\eta,\frac{1}{\sqrt2}
\right)
\left[-d\eta^{2}+dx^{2}+dy^{2}+dz^{2}\right]. \label{eq:ncmetric}
\end{align}
\\
The constant of motion for eq \bref{eq:ncmetric} is given by 

\begin{align}
\mathfrak{C} = \frac{\Lambda}{3}, 
\end{align}
\\
which then sets the energy density $\rho_0$ to
\\
\begin{align}
\rho_{0} = -\frac{\Lambda}{8\pi G_N \left(\mathrm{nc}\left(\omega \eta,\frac{1}{\sqrt{2}}\right )\right)^4},  \label{eq:ncdefrhoval}
\end{align}
\\
which is always negative for $\Lambda > 0$.\\
The squared scale factor $\phi(t)$  in cosmic time for eq \bref{eq:ncmetric} is 

\begin{align}
\phi^2(t)
=
\cosh\!\left[
2\sqrt{\frac{\Lambda}{3}}(t-t_0)
\right],
\end{align}
\\
where $t_{0}$ is the time of minimum expansion for the universe, with scale factor $\phi(t_0) = 1$. This solution corresponds to a spacetime which is infinitely expanded at $t<t_0$, contracts, then reaches some minimum expansions at $t_0$, then infinitely expands for $t > t_0$.
\\
Again, the results above for the choice $\phi_{3}$ are related to the results that would follow from the choice of $\phi_{5} = \mathrm{ds}$, once one recalls the identity
\\
\begin{align}
\mathrm{ds}(u,k)
= -\frac{1}{\sqrt2}\,\mathrm{nc}(u+K,k),
\end{align}
\\
with $\omega$ shifted as follows

\begin{align}
\omega \rightarrow \frac{\sqrt{2}}{2} \omega
\end{align} 
\\
which once more demonstrates, at least in spirit, that one can obtain (up to a common phase and scaling) the result for $\phi_{5}$ by numerically adjusting $\phi_3$.

\subsection{Electrovacuum Solutions}
\label{sec:electrovac}

\begin{figure}[h!]
\centering
\includegraphics[width=0.8\textwidth]{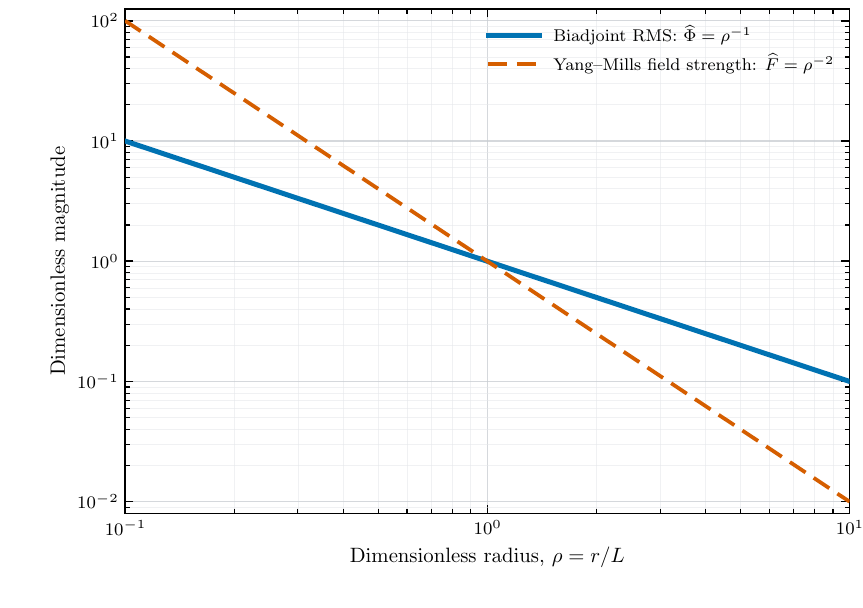}
\caption{A log--log comparison of the normalised radial profiles of the quartic
biadjoint scalar and Yang--Mills field strength. In terms of the
dimensionless radius $\rho=r/L$, the scalar amplitude scales as
$\widehat{\Phi}=\rho^{-1}$, whereas the field-strength magnitude scales
as $\widehat{F}=\rho^{-2}$. The normalisations are chosen such that both
profiles equal unity at $\rho=1$, and the singular point $\rho=0$ is
excluded.}
\label{fig:field-profile-falloff}
\end{figure} 

The Bertotti--Robinson spacetime \cite{bertotti1959uniform,Robinson:1959ev} describes a spacetime geometry which is filled with a constant electromagnetic field which can be described geometrically as the direct product of \(\mathrm{AdS}_2 \times S^2\), for cosmological constant $\Lambda = 0$.
Writing the flat metric in spherical coordinates, it takes the manifestly conformal form \cite{ottewill2012quantum,gron2013different}
\\

\begin{align}
ds^{2}
=\frac{L^{2}}{r^{2}}
\left(-dt^{2}+dr^{2}+r^{2}d\Omega_{2}^{2}\right)
=ds^{2}_{AdS_{2}(L)}+ds^{2}_{S^{2}(L)}, 
\label{eq:bertottirobinsonmetric}
\end{align}
\\
with $L$ as the radius of the space ${S^{2}(L)}$ and $\phi$ given as\footnote{$r$ takes the standard definition $r =\sqrt{x^2 + y^2 +z^2}$.}

\begin{align}
\phi=\frac{L}{r}. 
\end{align} 
\\
From the line element given in eq \bref{eq:bertottirobinsonmetric} yields the Ricci tensor 

\begin{align}
R_{\mu\nu}=
\begin{pmatrix}
\frac{1}{r^2} & 0 & 0 & 0 \\
0 & -\frac{1}{r^2} & 0 & 0 \\
0 & 0 & 1 & 0 \\
0 & 0 & 0 & \sin^2\theta
\end{pmatrix}
\label{eq:bertottirobinsonricci}
\end{align}
\\
which \cite{Torre:2013nia,doi:10.1073/pnas.10.4.124} obeys the Rainich conditions for solutions of the Einstein-Maxwell equations. 
Remembering that $\mathcal{T}_{\mu\nu}$ for the Einstein-Maxwell equations takes the form (for some dimensional parameter $\mu_{0}$)

\begin{align}
\mathcal{T}_{\mu\nu} =
\frac{1}{{\mu_0}}
\left(
F_{\mu\alpha}F_{\nu}{}^{\alpha}
-\frac14 g_{\mu\nu}F_{\alpha\beta}F^{\alpha\beta}
\right), \label{eq:brEMT}
\end{align}
\\
 we can then derive the electromagnetic field strength tensor which sources the energy-momentum tensor $\mathcal{T}_{\mu\nu}$
 \begin{align}
 F_{\mu \nu} =
\begin{pmatrix}
0 & \frac{Q}{r^2} & 0 & 0 \\
-\frac{Q}{r^2} & 0 & 0 & 0 \\
0 & 0 & 0 & 0 \\
0 & 0 & 0 & 0
\end{pmatrix}.
\end{align}
\\
where the inclusion of some charge $Q$ imposes that 
\begin{align}
Q^2=\frac{\mu_0 L^2}{4\pi G_N},
\end{align}
\\
The corresponding Yang-Mills gauge field and quartic biadjoint scalar field are
\begin{align}
A_{\mu} = -i \sigma_{\mu \nu} \partial^{\nu}  \log \left( r\right), \label{eq:GaugeBR}
\end{align}

\begin{align}
\Phi^{a a'} = \mu U^{a a'} \frac{L}{r}.
\end{align}
\\
Figure \ref{fig:field-profile-falloff} compares the relative strengths of
the biadjoint field and the Yang--Mills gauge field. The biadjoint field
is proportional to $r^{-1}$, whereas the field strength is proportional to
$r^{-2}$.

\section{Comparison to the Kerr-Schild Double Copy}
\label{sec:classical}
As discussed in the introduction, the Kerr-Schild double copy provided the first classical double copy in the literature, relating Kerr-Schild solutions in gravity to those in electromagnetism. Explicitly, so-called generalised Kerr-Schild Solutions in gravity can have their metrics written as 

\begin{align}
g_{\mu\nu}
=
\bar g_{\mu\nu}
+ \kappa^{2}
\phi\,k_\mu k_\nu.
\label{eq:genks-metric}
\end{align}
\\
here $\bar g_{\mu\nu}$ is the generalised background metric,
$\kappa^2$ is the gravitational coupling constant, $\phi$ is some scalar field and $k_{\mu}$ is a Kerr-Schild vector, satisfying the null and geodesic condition 

\begin{align}
\bar g^{\mu\nu}k_\mu k_\nu = 0,
\qquad
k^\nu \bar{\nabla}_\nu k^\mu = 0.
\label{eq:ks-conditionsgen}
\end{align} 
\\
Via the double copy (or more specifically the single copy operation), eq \eqref{eq:genks-metric} can be mapped to electromagnetic solutions with gauge fields $a_{\mu}$ of the form 

\begin{align}
a_{\mu} = k_{\mu} \phi. \label{eq:ksemgauge}
\end{align} 
\\
Using our previous results derived in section \ref{sec:electrovac} for the Bertotti--Robinson solution, we can derive a new single copy interpretation of the metric defined in eq \bref{eq:genks-metric}. \\
\\
Firstly, we can simplify the gauge field counterpart for the Bertotti--Robinson solution by evaluating the partial derivative in eq \bref{eq:GaugeBR}\footnote{The introduction of the bar above the gauge field in eq \bref{eq:GaugeBR2} is purely symbolic, and is only meant to distinguish it as the background gauge field rather than the full gauge field we wish to consider in eq \bref{eq:br-type-a-ansatz}.}

\begin{align}
\bar{A}_{\mu} = -\ii \sigma_{\mu i} \frac{n^{i}}{r}, \label{eq:GaugeBR2}
\end{align}
\\
with $n^{i}$ defined as 

\begin{align}
n^i\equiv \frac{x^i}{r},
\end{align}
\\
for $x^i\in\{x,y,z\}$. The gauge field can be decomposed into purely
temporal and spatial components\footnote{$n_{i} \equiv n^{i}$.}
\begin{align}
&\bar{A}_{0} = -\frac{\sigma_{i} n^{i}}{2r} \label{eq:BRTIME} \\ 
&\bar{A}_{i} = -\frac{\ii}{2r} \epsilon_{ijk}n_{j}\sigma_{k} \label{eq:BRSPACE}
\end{align}
\\
Comparing equations \bref{eq:BRTIME} and \bref{eq:BRSPACE} with eq \bref{eq:colourgaugedef} then yields 

\begin{align}
\bar A_0^a
&=
-\frac{\ii}{g r}\,n^a,
&
\bar A_i^a
&=
\frac{1}{g r}\,\epsilon_{aij}n^j,
\label{eq:br-ym-components}
\end{align}
\\
whose field strength tensor $\bar{F}_{\mu \nu}^{a}$ satisfies the field equation 

\begin{align}
\bar{D}_{\mu} \bar{F}^{\mu \nu a} = 0, 
\end{align}
where $\bar{D}_{\mu}$ is the covariant derivative associated with the field strength tensor $\bar{F}_{\mu\nu}^{a}$.
 \\ 
A key observation is that, by construction $n_{i}$ satisfies 
\begin{align}
\bar{D}_{\mu} n_{i} = 0
\end{align}
\\
which then allows us to build a modified gauge potential $A_{\mu}^{a}$
from an Abelian gauge potential $a_\mu$ and $\bar{A}_{\mu}^{a}$
\begin{align}
A_\mu^a
=
\bar A_\mu^a+\frac{1}{g}n^a a_\mu, 
\label{eq:br-type-a-ansatz}
\end{align}
\\
with field strength $F_{\mu \nu}^{a}$ 

\begin{align}
F_{\mu\nu}^a
&=
\bar F_{\mu\nu}^a+\frac{1}{g}n^a f_{\mu\nu}. \label{eq:BRfamend}
\end{align}
\\
Equation \bref{eq:BRfamend} then satisfies the equation of motion 

\begin{align}
\left(D_\mu F^{\mu\nu}\right)^a
&=
\frac{1}{g}n^a\partial_\mu f^{\mu\nu}, 
\label{eq:br-type-a-reduction}
\end{align} 
\\
Given that $f_{\mu\nu}$ satisfies the Abelian equations of motion, eq
\bref{eq:BRfamend} implies that $F^a_{\mu\nu}$ satisfies the Yang--Mills
equations of motion, i.e.,
 \begin{align}
\left(D_\mu F^{\mu\nu}\right)^a
&= 0.
\label{eq:br-type-a-reduction-zero}
\end{align} 
\\
Consequently, equation \bref{eq:br-type-a-reduction-zero} then determines that any allowed choice of $a_{\mu}$ maybe be added to $\bar{A}_{\mu}^{a}$ to yield a full solution to the SU(2) Yang-Mills equation. Of interest to the classical double copy, this allows us to choose, if we so wished 

\begin{align}
a_{\mu} = \psi k_{\mu}, \label{eq:smallaks}
\end{align} 
\\
i.e the Kerr-Schild ansatz for the single copy in eq \bref{eq:ksemgauge}, for some scalar field $\psi$.
\\
The above construction therefore allow us to determine an interesting classical double copy relationship for the Bertotti--Robinson solution $\bar{g}^{BR}_{\mu \nu}$, schematically, as

\begin{align}
\begin{array}{c}
g_{\mu\nu}
=
\bar{g}_{\mu\nu}^{BR}
+
\psi\,k_{\mu}k_{\nu}
\\[1.2ex]
\text{Single copy}\;\downarrow
\qquad
\uparrow\;\text{Double copy}
\\[1.2ex]
A_{\mu}^{a}
=
\bar{A}_{\mu}^{a}
+
\dfrac{n^{a}}{g}\,\psi k_{\mu}
\end{array}
\end{align}
\\
with explicit identifications of 

\begin{align}
\bar{g}_{\mu\nu}^{BR} \rightarrow \bar{A}^{a}_{\mu}, \\
\psi\,k_{\mu}k_{\nu} \rightarrow \dfrac{n^{a}}{g}\,\psi k_{\mu}.
\end{align}
\\
Ultimately, although the solution in eq. \bref{eq:br-type-a-reduction} is
a full $SU(2)$ solution, it reduces the effective dynamics to a $U(1)$
Abelian subsector. Thus, the construction should not itself be viewed as a
fully nonlinear relationship.

\section{Discussion and Further Work}
\label{sec:discussion}

In this paper, we have constructed an exact double copy correspondence between restricted sectors of quartic biadjoint scalar theory, $SU(2)$ Yang--Mills theory and four-dimensional general relativity.  The three ansatz fields are generated by a common scalar profile $\phi$ and satisfy their respective equations of motion provided that
\begin{equation}
  4\mu^2\lambda_4=\lambda=\frac{2\Lambda}{3}.
\end{equation} 
where $4\mu^2\lambda_4$, $\lambda$ and $\frac{2\Lambda}{3}$ are the coupling constants for the reduced scalar equations for quartic biadjoint scalar theory, Yang-Mills theory, and gravity, respectively. In the Yang-Mills case, $\lambda$ is an arbitrary integration constant with mass dimension $2$ and $\Lambda$ (also with mass dimension 2) is the cosmological constant in gravity.
The resulting equation, $\Box\phi+\lambda\phi^3=0$, retains a genuine nonlinearity in each of the three reduced descriptions.  In particular, the quartic biadjoint interaction is essential: after the factorised colour ansatz it produces the same cubic term in $\phi$ as the Yang--Mills and gravitational reductions.  
\\
\\
A central part of the dictionary is the treatment of sources.  The scalar equation follows only from the trace of the Einstein equations and is not, by itself, sufficient to determine a gravitational solution.  Retaining the trace-free equation fixes the admissible traceless source and on the interacting branch $\lambda\neq0$, yields the conformal relation between the energy momentum tensors for the Yang-Mills sector and Gravity sector. Using these results, we can then reinterpret this previously known conformal relationship between energy momentum tensors as signifying an actual non-perturbative classical double copy   
\\
\\
We then discussed applicable examples of this new non-perturbative correspondence. Rational and constant profiles reproduce $AdS_4$, planar $dS_4$ and Minkowski space, while the Jacobi-elliptic profiles yield spatially flat radiation cosmologies.  The $\operatorname{cn}/\operatorname{sd}$ branch has positive energy density and recollapses for $\Lambda<0$, whereas the $\operatorname{nc}/\operatorname{ds}$ branch bounces for $\Lambda>0$ at the price of a negative radiation energy density.  The Bertotti--Robinson example is qualitatively different: its radial profile shows that the construction is outside of the original single variable parametrisation for $\phi$, for the harmonic (and in the case of Yang-Mills, self-dual) branch $\lambda=\Lambda=0$.
\\
\\
We then compared our results with the Kerr--Schild double copy described
in \cite{Monteiro:2014cda,Luna:2018dpt}. First, this comparison led us to
re-derive, from a novel perspective within nonlinear correspondences, the
gauge-theory counterpart of Minkowski space: a trivial gauge field
($A_\mu=0$).
Second, our results allowed us to derive a classical double copy for the
Kerr--Schild ansatz about the Bertotti--Robinson metric. This yielded, for
the first time, a background-plus-graviton correspondence between gravity
and solutions of the full $SU(2)$ Yang--Mills equations. This example is
not fully nonlinear, because the dynamics are effectively restricted to a
$U(1)$ subsector of the full $SU(2)$ theory.
\\
\\
Further work should focus on studying the origins of this nonlinear
correspondence more deeply and on exploring possible extensions of its
scope.
\\
\\
In regards to the latter, an embedding of the present solution into an $SU(2)$ subgroup of $SU(3)$ or $SU(N)$ is immediate, but introduces no genuinely new colour dynamics.  A nontrivial extension would require constant colour tensors and spacetime matrices whose contractions reduce the Yang--Mills and biadjoint equations to the same scalar equation without confining the fields to an $SU(2)$ subalgebra.  Classifying such tensors would show whether the role played here by the three-dimensional Levi--Civita symbol and the $O(3)$ matrix $U^{aa'}$ is accidental or part of a broader algebraic mechanism. One plausible route to perform this extension is to study relatively new non-linear solutions of classical SU(3) Yang Mills theory, like those seen in reference \cite{Tsapalis:2016fft}.
\\
\\
The best route to understanding the origin of this new relationship may be
to follow the methodology used to elucidate the original Kerr--Schild
double copy. More concretely, this would first mean exploring the
correspondence in the language of two-component spinors. As discussed in
the introduction, the compatible gravitational metrics under this ansatz
all have a vanishing Weyl tensor and thus cannot be interpreted using the
standard Weyl double copy \cite{Godazgar:2020zbv}. Nevertheless, there are
other spinorial quantities that could be explored instead of the Weyl
tensor. This is motivated by the fact that the spinorial translation of the
Riemann curvature tensor contains fields other than the Weyl tensor, such
as the Ricci tensor and cosmological term, so these structures could be
used when making a connection to gauge theory \cite{penrose1984spinors}.
Furthermore, the natural next step after studying the spinorial
representation of the correspondence would be to investigate twistor
methods such as those used in ref. \cite{Chacon:2021wbr}.\\
\\
Finally, the $AdS_4$ representative raises a possible holographic question.  The existence of an anti-de Sitter member of the solution family does not by itself define an AdS/CFT dictionary \cite{Maldacena:1997re,Witten:1998qj}.  To make such a connection precise, one would need to specify admissible boundary conditions, determine the boundary data induced by the scalar and Yang--Mills representatives, and compare renormalised observables rather than only bulk profiles.  Establishing such a boundary map, together with a genuinely non-Abelian extension of the Bertotti--Robinson background construction, would help determine whether the common scalar seed reflects a deeper double-copy structure or a special feature of the present ans\"atze.
\\
\\
We look forward to reporting back on the above and other directions in exploring this new non-perturbative relationship in the future.

\appendix

\acknowledgments

We would like to thank Mariana Carrillo González, Donal O'Connell, Andr\'es Luna, Nathan Moynihan, Chris D. White, for various discussions and feedback on the manuscript. KAW is supported by a Royal Society Career Development Fellowship.

\bibliographystyle{JHEP}
\bibliography{references.bib}


%
%
%



\end{document}